\documentclass[english]{elsarticle}
\usepackage[T1]{fontenc}
\usepackage[latin9]{inputenc}
\usepackage{url}
\usepackage{graphicx}
\usepackage{setspace}
\usepackage{babel}
\begin{document}
\begin{frontmatter}

\title{Domain-Specific Acceleration and Auto-Parallelization of Legacy Scientific
Code in FORTRAN 77 using Source-to-Source Compilation}

\author[gla]{Wim Vanderbauwhede}
\ead{wim.vanderbauwhede@glasgow.ac.uk}
\author[gla]{Gavin Davidson}
\ead{gavinaadavidson@gmail.com}
\address[gla]{School of Computing Science, University of Glasgow, Glasgow, Scotland}

\begin{abstract}
Massively parallel accelerators such as GPGPUs, manycores and FPGAs
represent a powerful and affordable tool for scientists who look to
speed up simulations of complex systems. However, porting code to
such devices requires a detailed understanding of heterogeneous programming
tools and effective strategies for parallelization. In this paper
we present a source to source compilation approach with whole-program
analysis to automatically transform single-threaded FORTRAN 77 legacy
code into OpenCL-accelerated programs with parallelized kernels. 

The main contributions of our work are: (1) whole-source refactoring
to allow any subroutine in the code to be offloaded to an accelerator.
(2) Minimization of the data transfer between the host and the accelerator
by eliminating redundant transfers. (3) Pragmatic auto-parallelization
of the code to be offloaded to the accelerator by identification of
parallelizable \emph{maps} and \emph{reductions}. 

We have validated the code transformation performance of the compiler
on the NIST FORTRAN 78 test suite and several real-world codes: the
Large Eddy Simulator for Urban Flows, a high-resolution turbulent
flow model; the shallow water component of the ocean model Gmodel;
the Linear Baroclinic Model, an atmospheric climate model and Flexpart-WRF,
a particle dispersion simulator. 

The automatic parallelization component has been tested on as 2-D
Shallow Water model (2DSW) and on the Large Eddy Simulator for Urban
Flows (UFLES) and produces a complete OpenCL-enabled code base. The
fully OpenCL-accelerated versions of the 2DSW and the UFLES are resp.
9x and 20x faster on GPU than the original code on CPU, in both cases
this is the same performance as manually ported code.
\end{abstract}
\end{frontmatter}

\section{Background}

\begin{figure}
\centering{}\includegraphics[width=0.6\textwidth]{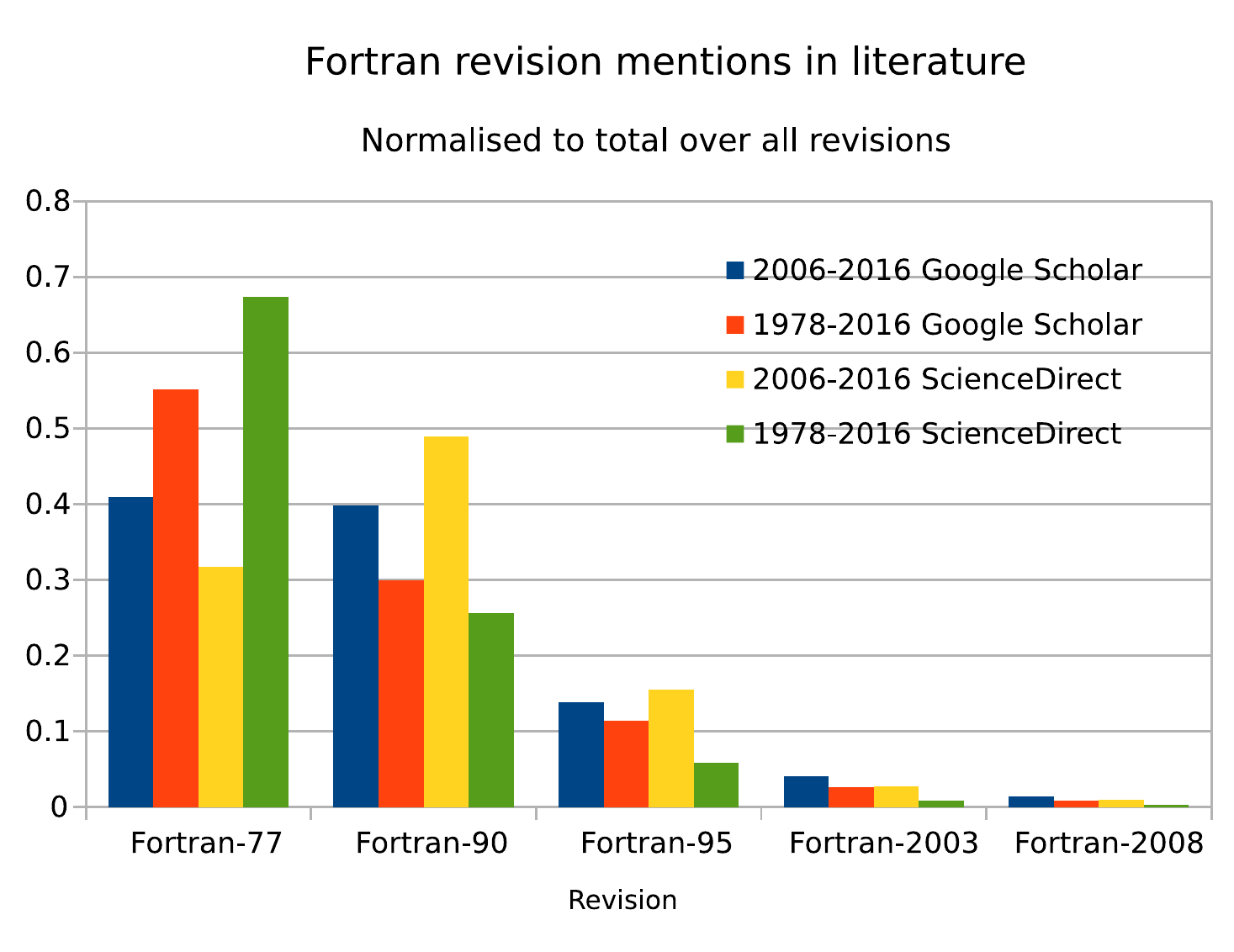}\caption{\label{fig:Literature-mentions-of}Literature mentions of different
revisions of Fortran using Google Scholar and ScienceDirect}
\end{figure}

A large amount of scientific code (both ``legacy'' code and new
code) is still effectively written in FORTRAN 77. Fig. \ref{fig:Literature-mentions-of}
shows the relative citations (citations per revision normalized to
sum of citations for all revisions) for Google Scholar and ScienceDirect
for each of the main revisions of Fortran. We collected results for
the past 10 years (2006-2016) and also since the release of FORTRAN
77 (1978-2016). As an absolute reference, there were 15,700 citations
in Google Scholar mentioning FORTRAN 77 between 2006 and 2016. It
is clear that FORTRAN 77 is still widely used and that the latest
standards (2003, 2008) have not yet found widespread adoption.

Based on the above evidence \textendash{} and also on our own experience
of collaboration with scientists \textendash{} the current state of
affairs is that for many scientists, FORTRAN 77 is still the language
of choice for writing models. There is also a vast amount of legacy
code in FORTRAN 77. Because the FORTRAN 77 language was designed with
assumptions and requirements very different from today's, code written
in it has inherent issues with readability, scalability, maintainability
and parallelization. A comprehensive discussion of the issues can
be found in \cite{tinetti2012fortran}. As a result, many efforts
have been aimed at refactoring legacy code, either interactive or
automatic, and to address one or several of these issues. 

Our work is part of that effort, but we are specifically interested
in automatically \emph{refactoring Fortran for OpenCL-based accelerators}.
In this paper we present a source compilation approach to transform
sequential FORTRAN 77 legacy code into high-performance OpenCL-accelerated
programs with auto-parallelized kernels without need for directives
or extra information from the user.

\section{Heterogeneous Computing and Accelerators\label{subsec:Heterogenenous-Computing-and}}

by heterogeneous computing we mean computing on a system comprising
a (multicore) host processor and an accelerator, e.g. a GPGPU, FPGA
or a manycore device such as the Intel Xeon Phi. Many scientific codes
have already been investigated for and ported manually to GPUs, and
excellent performance benefits have been reported. There are many
approaches to programming accelerators, but we restrict our discussion
to open standards and do not discuss commercial solutions tied to
a particular vendor or platform; and we will only discuss solutions
that work in Fortran.

\subsection{OpenCL}

The OpenCL framework\cite{stone2010opencl} presents an abstraction
of the accelerator hardware based on the concept of \emph{host} and
\emph{device}. A programmer writes one or more \emph{kernels} that
are run directly by the accelerator and a \emph{host program} that
is run on the system's main CPU. The host program handles memory transfers
to the device and initializing computations and the kernels do the
bulk of the processing, in parallel on the device. The main advantage
of OpenCL over proprietary solutions such as e.g. CUDA (to which it
is very similar) is that it supported by a wide range of devices,
including multicore CPUs, FPGAs and GPUs. From the programmer perspective,
OpenCL is very flexible but quite low level and requires a lot of
boilerplate code to be written. This is a considerable barrier for
adoption by scientists. Furthermore, there is no official Fortran
support for OpenCL: the host API is C/C++, the kernel language is
based on a subset of C99. To remedy this we have developed \cite{vanderbauwhede2015twinned}
a Fortran API for OpenCL\footnote{https://github.com/wimvanderbauwhede/OpenCLIntegration}.

\subsection{OpenACC and OpenMP}

OpenACC\footnote{https://www.openacc.org/} takes a directive based
approach to heterogeneous programming that affords a higher level
of abstraction for parallel programming than OpenCL or CUDA. In a
basic example, a programmer adds \emph{pragmas} (compiler directives)
to the original (sequential) code to indicate which parts of the code
are to be accelerated. The new source code, including directives,
is then processed by the OpenACC compiler and programs that can run
on accelerators are produced. There are a number of extra directives
that allow for optimization and tuning to allow for the best possible
performance.

With OpenMP version 4, the popular OpenMP standard\footnote{http://www.openmp.org/}
for shared-memory parallel programming now also supports accelerators.
The focus of both standards is slightly different, the main difference
being that OpenMP allows conventional OpenMP directives to be combined
with accelerator directives, whereas OpenACC directives are specifically
designed for offloading computation to accelerators.

Both these annotation-based approaches are local: they deal with parallelization
of relatively small blocks and are not aware of the whole code base,
and this makes them both harder to use and less efficient. To use
either on legacy FORTRAN 77 code, it is not enough to insert the pragmas:
the programmer has to ensure that the code to be offloaded is free
of global variables, which means complete removal of all common block
variables or providing a list of shared variables as annotation. The
programmer must also think carefully about the data movement between
the host and the device, otherwise performance is poor. 

\subsection{Raising the abstraction level}

Our approach allows an even higher level of abstraction than that
offered by OpenACC or OpenMP: the programmer does not need to consider
how to achieve program parallelization, but only to mark (using a
single annotation) which subroutines will be paralleled and offloaded
to the accelerator. Our compiler provides a fully automatic conversion
of a complete FORTRAN 77 codebase to Fortran 95 with OpenCL kernels.
Consequently, the scientists can keep writing the code in FORTRAN
77, and the original code base is always intact. 

\section{Existing source-to-source compilers and refactoring tools}

A conventional compiler consumes source code and produces binaries.
A source-to-source compiler produces transformed source code from
the original source. This transformation can be e.g. refactoring,
parallelization or translation to a different language. The advantage
is that the resulting code can be modified by the programmer if desired
and compiled with a compiler of choice. 

There are a number of source-to-source compilers and refactoring tools
for Fortran available. However, very few of them actually support
FORTRAN 77. The most well known are the ROSE framework\footnote{\url{http://www.rosecompiler.org/index.html}}
from LLNL \cite{liao2010rose}, which relies on the Open Fortran Parser
(OFP)\footnote{\url{http://fortran-parser.sourceforge.net/}} . This
parser claims to support the Fortran 2008 standard. Furthermore, there
is the language-fortran\footnote{\url{https://hackage.haskell.org/package/language-fortran}}
parser which claims to support FORTRAN 77 to Fortran 2003. A  refactoring
framework which claims to support FORTRAN 77 is CamFort \cite{Orchard:2013:UFS:2541348.2541356},
according to its documentation it supports Fortran 66, 77, and 90
with various legacy extensions.

We tested OFP 0.8.3, language-fortran 0.5.1 and CamFort 0.804  using
the NIST FORTRAN 78 test suite (discussed in more detail in Section
\ref{sec:Validation}). All three parsers failed to parse any of the
provided sources. Consequently we could not use either of these as
a starting point.

Like CamFort, the Eclipse-based interactive refactoring tool Photran
\cite{overbey2005refactorings}, which supports FORTRAN 77 - 2008,
is not a whole-source compiler, but works on a per-file basis (which
is in fact what most compilers do). Both CamFort and Photran provide
very useful refactorings, but these are limited to the scope of a
code unit. For effective refactoring of common blocks, and determination
of data movement direction, as well as for effective acceleration,
whole-source code (inter-procedural) analysis and refactoring is essential. 

A long-running project which does support inter-procedural analysis
is PIPS\footnote{\url{http://pips4u.org/}}, started in the 1990's.
The PIPS tool does support FORTRAN 77 but does not supported the refactorings
we propose. Support for autoparallelization via OpenCL was promised
\cite{amini2011pips} but has not yet materialized. For completeness
we mention the commercial solutions plusFort \footnote{\url{http://www.polyhedron.com/pf-plusfort0html}}
and VAST/77to90 \footnote{\url{http://www.crescentbaysoftware.com/compilertech.html}}
which both can refactor common blocks into modules but not into procedure
arguments.

\section{Our Goal and Approach}

FORTRAN 77 code is often computationally efficient, and programmer
efficient in terms of allowing the programmer to quickly write code
and not be too strict about it. As a result it becomes very difficult
to for maintain and port. Our goal is that the refactored code should
meet the following requirements:

\subsection{Modern, Maintainable and Extensible }

FORTRAN 77 was designed with very different requirements from today's
languages, notably in terms of avoiding bugs. It is said that C gives
you enough rope to hang yourself. If that is so then FORTRAN 77 provides
the scaffold as well. Specific features that are unacceptable in a
modern language are:
\begin{itemize}
\item Implicit typing, i.e. an undeclared variable gets a type based on
its starting letter. This may be very convenient for the programmer
but makes the program very hard to debug and maintain. Our compiler
makes all types explicit (\texttt{implicit none}).
\item No indication of the intended access of subroutine arguments: in FORTRAN
77 it is not possible to tell if an argument will be used read-only,
write-only or read-write. This is again problematic for debugging
and maintenance of code. Our compiler infers the \texttt{intent} for
all subroutine and function arguments. 
\item In FORTRAN 77, procedures defined in a different source file are not
identified as such. For extensibility as well as for maintainability,
a module system is essential. Our compiler converts all non-program
code units into modules which are \texttt{use}d with an explicit export
(\texttt{only}) declaration. 
\end{itemize}
There are several more refactorings that our compiler applies, such
as rewriting label-bases loops as do-loops etc, but they are less
important for this paper.

\subsection{Accelerator-ready}

As discussed in Section \ref{subsec:Heterogenenous-Computing-and},
the common feature of the vast majority of current accelerators is
that they have a separate memory space, usually physically separate
from the host memory. Furthermore, the common offload model is to
create a ``kernel'' subroutine (either explicitly or implicitly)
which is run on the accelerator device. Consequently, it is crucial
to separate the memory spaces of the kernel and the host program. 
\begin{itemize}
\item FORTRAN 77 programs makes liberal use of global variables through
``\texttt{common}'' blocks. Our compiler converts these common block
variables into subroutine arguments across the \emph{complete} call
tree of the program. Although refactoring of common blocks has been
reported for some of the other projects, to our knowledge our compiler
is the first to perform this refactoring across multiple nested procedure
calls, potentially in different source code units. 
\end{itemize}

\subsection{Automatic Parallelization and Acceleration}

Our ultimate goal is to convert legacy FORTRAN 77 code into parallel
code so that the computation can be accelerated using OpenCL. We
use a three-step process:

First, the above refactorings\footnote{https://github.com/wimvanderbauwhede/RefactorF4ACC}
result in a modern, maintainable, extensible and accelerator-ready
Fortran 95 codebase. This is an excellent starting point for many
of the other existing tools, for example the generated code can now
easily be paralleled using OpenMP or OpenACC annotations, or further
refactored if required using e.g. Photran or PIPS. However, we want
to provide the user with an end-to-end solution that does not require
any annotations. 

The second step in our process is to identify data-level parallelism
present in the code in the form of \emph{maps} and \emph{folds} (i.e.
loops without dependencies and reductions). The terms \emph{map} and
\emph{fold} are taken from functional programming and refer to ways
of performing a given operation on all elements of a list. Broadly
speaking these constructs are equivalent to loop nests with and without
dependencies, and as Fortran is loop-based, our analysis in indeed
an analysis of loops and dependencies. However, our internal representation
uses the functional programming model where \emph{map} and \emph{fold}
are functions operating on other functions (i.e. they are higher-order
functions), the latter being extracted from the bodies of the loops.
Thus we raise the abstraction level of our representation and make
it independent of both the original code and the final code to be
generated. We apply a number of rewrite rules for map- and fold-based
functional programs (broadly speaking equivalent to loop fusion or
fission) to optimist the code.

The third step is to generate OpenCL host and device code from the
paralleled code. Because of the high abstraction level of our internal
representation, we could easily generate OpenMP or OpenACC annotations,
CUDA or Maxeler's MaxJ language used to program FPGAs. Our compiler\footnote{https://github.com/wimvanderbauwhede/AutoParallel-Fortran}
also minimizes the data transfer between the host and the accelerator
by eliminating redundant transfers. This includes determining which
transfers need to be made only once in the run of the program.

\section{Code Transformation Validation\label{sec:Validation}}

To assess the correctness and capability of our refactoring compiler,
we used the NIST (US National Institute of Standards and Technology)
FORTRAN 78 test suite \footnote{http://www.itl.nist.gov/div897/ctg/fortran\_form.htm},
which aims to validate adherence to the ANSI X3.9-1978 (FORTRAN 77)
standard. We used a version with some minor changes\footnote{http://www.fortran-2000.com/ArnaudRecipes/fcvs21\_f95.html}
: All files are properly formed; a non standard conforming FORMAT
statement has been fixed in test file FM110.f; Hollerith strings in
FORMAT statements have been converted to quoted strings. This test
suite comprises about three thousand tests organized into 192 files.
We skipped a number of tests because they test features that our compiler
does not support. In particular, we skipped tests that use spaces
in variable names and keywords (3 files, 23 tests) and tests for corner
cases of common blocks and block data (2 files, 37+16 tests). After
skipping these types of tests, 2867 tests remain, in total 187 files
for which refactored code is generated. The test bench driver provided
in the archive skips another 8 tests because they relate to features
deleted in Fortran 95. In total the test suite contains 72,473 lines
of code (excluding comments). Two test files contain tests that fail
in gfortran 4.9 (3 tests in total).

Our compiler successfully generates refactored code for all tests,
and the refactored code compiles correctly and passes all tests (2864
tests in total).

Furthermore, we tested the compiler on a simple 2-D shallow water
model from \cite{kampf2009ocean} (188 loc) and on four real-word
simulation models: the Large Eddy Simulator for Urban Flows \footnote{https://github.com/wimvanderbauwhede/LES},
a high-resolution turbulent flow model\cite{vanderbauwhede2015twinned}
(1,391 loc); the shallow water component of Gmodel\footnote{http://www.sciamachy-validation.org/research/CKO/gmodel.html},
an ocean model\cite{burgers2002balanced} (1,533 loc); Flexpart-WRF\footnote{https://github.com/sajinh/flx\_wrf2},
a version of the Flexpart particle dispersion simulator\cite{brioude2013lagrangian}
that takes input data from WRF (13,829 loc); and the Linear Baroclinic
Model\footnote{http://ccsr.aori.u-tokyo.ac.jp/\textasciitilde{}hiro/sub/lbm.html},
an atmospheric climate model\cite{watanabe2003moist} (39,336 loc). 

Each of these models has a different coding style, specifically in
terms of the use of common blocks, include files, etc that affect
the refactoring process. All of these codes are refactored fully automatically
without changes to the original code and build and run correctly.
The performance of the original and refactored code is the same in
all cases.

\section{Automatic Parallelization Evaluation}

In this section we show the performance of the automatically generated
OpenCL code compared to the best achievable performance of the unmodified
original code. We show that the automatically generated OpenCL code
can perform as well as hand-ported OpenCL code.

\subsection{Experimental Setup}

To evaluate the automatic parallelization and OpenCL code generation
we used following experimental setup: the host platform is an Intel
Xeon CPU E5-2620@2.00GHz, a 6-core CPU with hyperthreading (12 threads),
AVX, 32GB RAM, and 15MB cache; the GPU is an NVIDIA GeForce GTX TITAN,
980 MHz, 15 compute units, 16GB RAM. We used OpenCL 1.1 via the CUDA
6.5.14 SDK. The original UFLES code on CPU (reference) was compiled
with gfortran 4.8.2 with following flags for auto-vectorization and
auto-parallelization:\texttt{ -Ofast -floop-parallelize-all -ftree-parallelize-loops=12
-fopenmp -pthread}. Auto-parallelization provides only 4\% speed-up
because the most time-consuming loops are not paralleled. Our compiler
auto-parallelizes all loop nests in the code base and produces a complete
OpenCL-enabled code base that runs on GPU and CPU.

\subsection{Test Case 1: 2-D Shallow Water Model}

As a first test case for the validation of our automatic parallelization
approach we used the 2-D Shallow Water model from the textbook \cite{kampf2009ocean}
by Kaempf. This very simple model consists of a time loop which calls
two subroutines, a predictor (\emph{dyn}) and a first-order Shapiro
filter (\emph{shapiro}), before updating the velocity. Our compiler
automatically transforms this code into three map-style kernels.

\begin{figure}
\begin{centering}
\includegraphics[width=8cm]{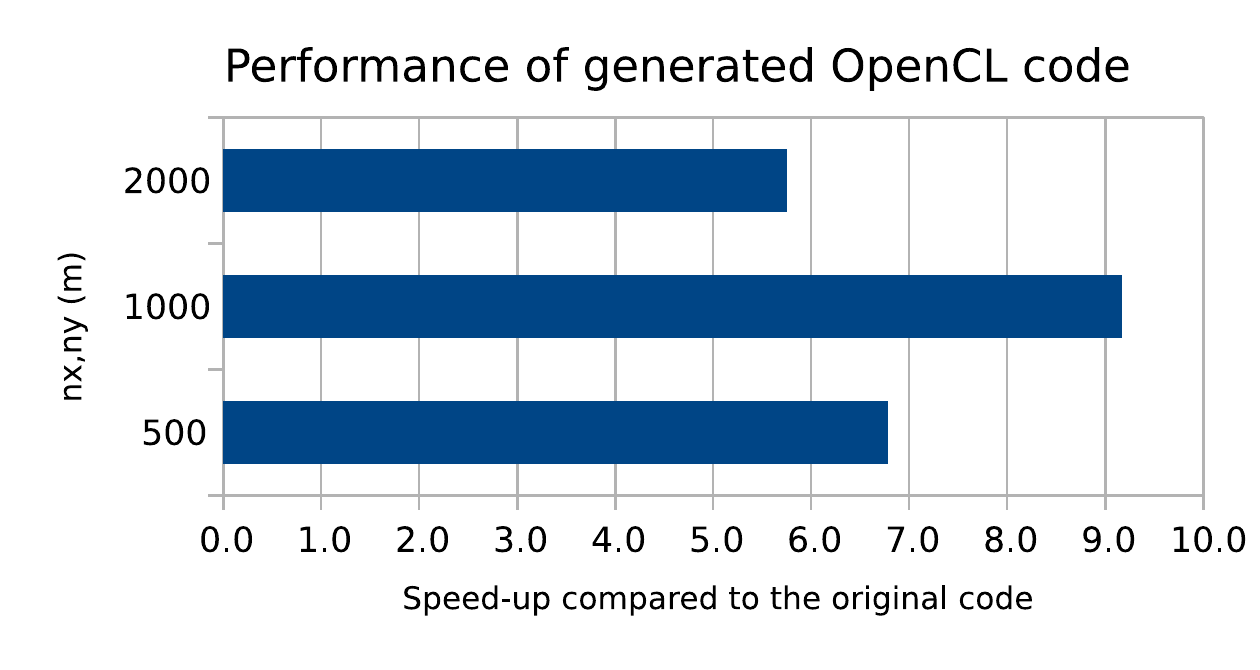}
\par\end{centering}
\caption{Speed-up compared to the original code\label{fig:Speed-up}}
\end{figure}

The results shown in Fig. \ref{fig:Speed-up} are for domain size
of 500x500,1000x1000, and 2000x2000 for 10,000 time steps. This is
a high-resolution simulation with spatial resolution of 1 m and a
time step of 0.01 s. The automatically generated code running on GPU
is up to 9x faster than the original code. This is the same performance
as obtained by manual porting of the code to OpenCL.

\subsection{Test Case 2: Large Eddy Simulator for Urban Flows (UFLES)}

As a more comprehensive test case we used the Large Eddy Simulator
for Urban Flows (UFLES) developed by Prof. Takemi at the Disaster
Prevention Research Institute of Kyoto University and Dr. Nakayama
of the Japan Atomic Energy Agency \cite{nakayama2011analysis}. This
simulator generates turbulent flows by using mesoscale meteorological
simulations. It explicitly represents the urban surface geometry using
GIS data and is used to conduct building-resolving large-eddy simulations
of boundary-layer flows over urban areas under realistic meteorological
conditions. The simulator essentially solves the Poisson equation
for the pressure using Successive Over-Relaxation and integrates the
force fields using the Adams-Bashforth algorithm.

\subsubsection{Functional Code Structure of UFLES}

The UFLES  main loop sequentially executes 7 subroutines consecutively
for each simulation time step:
\begin{description}
\begin{spacing}{0.5}
\item [{\textsf{velnw:}}] \textsf{Update velocity for current time step}
\item [{\textsf{bondv1:}}] \textsf{Calculate boundary conditions (initial
wind profile, inflow, outflow)}
\item [{\textsf{velfg:}}] \textsf{Calculate the body force }
\item [{\textsf{feedbf:}}] \textsf{Calculation of building effects (Goldstein
damping model)}
\item [{\textsf{les:}}] \textsf{Calculation of viscosity terms (Smagorinsky
model)}
\item [{\textsf{adam:}}] \textsf{Adams-Bashforth time integration}
\item [{press:}] Solving of Poisson equation using SOR (iterative solver)
\end{spacing}
\end{description}
Our compiler automatically transforms this code into 29 map-stle kernels
and 4 reduction kernels.

\subsubsection{OpenCL UFLES Results}

\begin{figure}[h]
\begin{centering}
\includegraphics[width=0.66\textwidth]{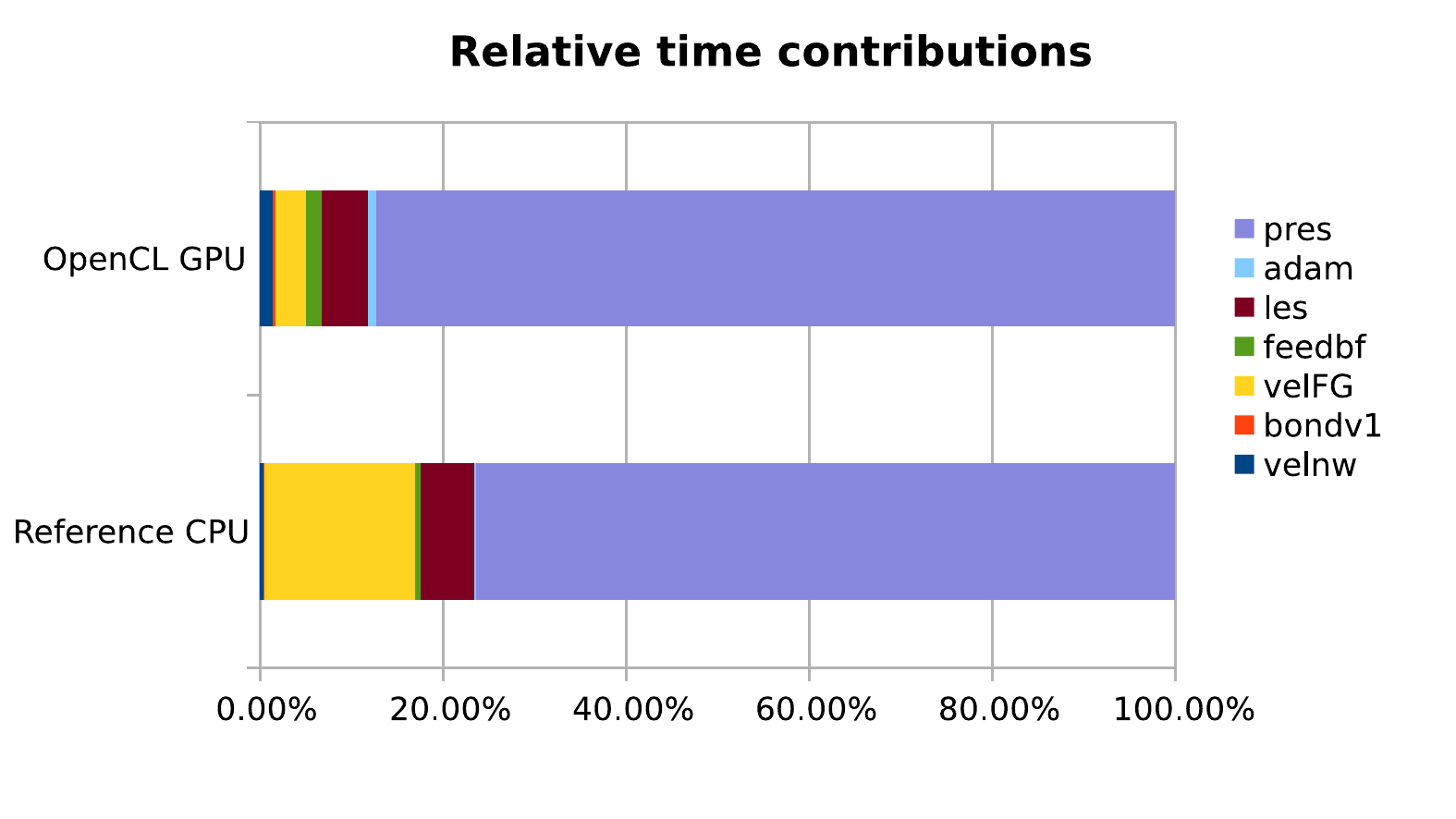}
\par\end{centering}
\centering{}\caption{Breakdown of time contribution per subroutine\label{fig:Breakdown-per-subroutine}}
\end{figure}

\begin{figure}

\begin{centering}
\includegraphics[width=0.66\textwidth]{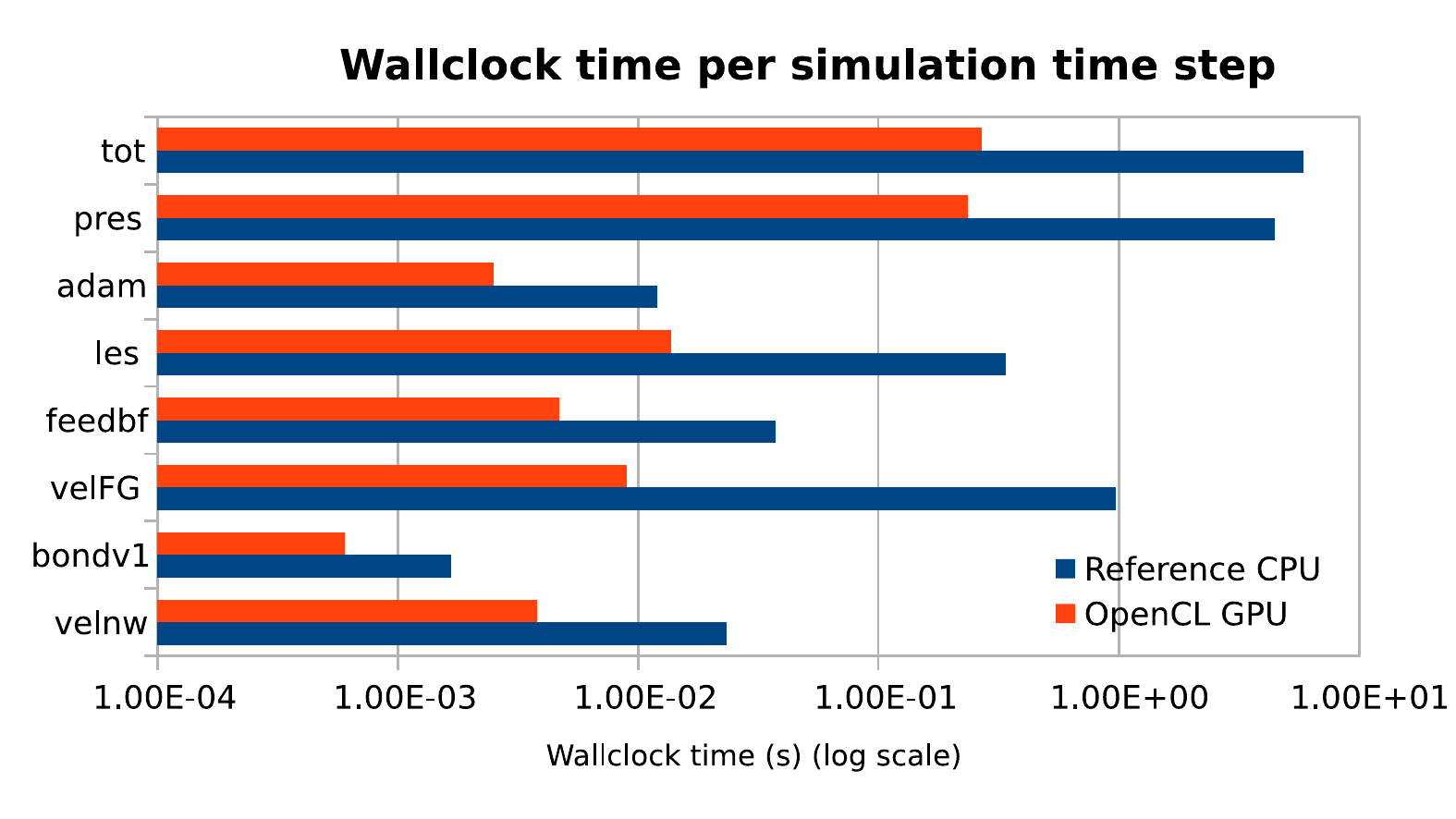}
\par\end{centering}
\caption{Wall clock time\label{fig:Wall-clock-time}}

\end{figure}

All results shown in Figs. \ref{fig:Breakdown-per-subroutine}, \ref{fig:Wall-clock-time}
and \ref{fig:Speed-up} are for a domain size of 300x300x90, with
the number of SOR iterations set to 50. This is a realistic use case
of the UFLES covering an area of 1.2km x 1.2km. A simulation time
step represents 0.025s of actual time. 

\begin{figure}[h]
\centering{}\includegraphics[width=0.66\textwidth]{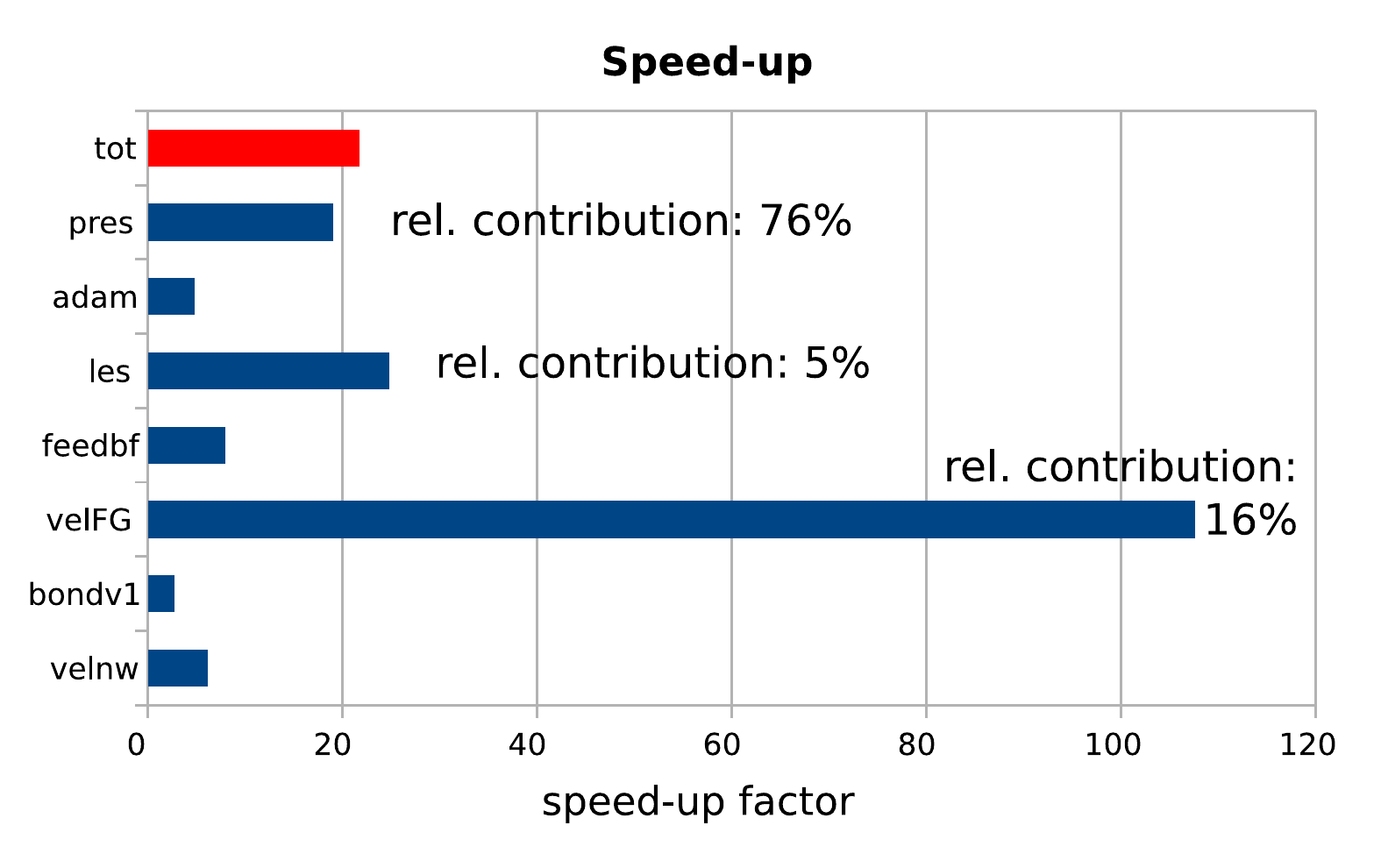}\caption{Speed-up compared to the original code\label{fig:Speed-up-compared-to}}
\end{figure}

Fig. \ref{fig:Breakdown-per-subroutine} shows the breakdown of relative
run time contributions per subroutine. We can see that the \emph{pres}
subroutine which contains the SOR iterative loop dominates the run
time. On the GPU, this routine accounts for almost 90\% of the run
time. Fig. \ref{fig:Wall-clock-time} shows the total wall clock time
and wall clock times for each subroutine on CPU and GPU. Note that
the scale is logarithmic. The main observations are that the GPU code
is faster for all subroutines but especially so for the \emph{velFG}
routine. Finally, Fig. \ref{fig:Speed-up-compared-to} shows the total
speed-up and the speed-up per subroutine. The speed-up of more than
100x for \emph{velFG} is remarkable. This is because this routine
performs a large amount of computations per point in the domain and
each point is independent. Thus the GPU can optimally exploit the
available parallelism. However, the total speed-up is entirely dominated
by the iterative SOR solver, which is 20x faster on the GPU. Our auto-parallelized
version achieves the same performance as the manually ported OpenCL
version of the UFLES \cite{vanderbauwhede2015twinned}.

\section{Discussion}

The above results demonstrate that it is possible to automatically
generate high-performance GPU code from FORTRAN 77 legacy code. All
the compiler expects the programmer to do is annotate a region of
the code for offloading. All subroutines in this region will be offloaded
to the accelerator. 

In practice there are some limitations. We have only presented two
examples because the autoparallelizing compiler currently lacks a
recursive inliner so that it only supports kernel subroutines that
do not call other subroutines.

We use the term ``domain specific'' not in the sense of a particular
branch of science but rather a of class of models: in essence, we
require the loop bounds to be static, i.e. known at compile time,
in order to parallels the loops. For the same reason, recursion is
not supported; however, recursion is not supported by the ANSI X3.9-1978
(FORTRAN 77 standard). Furthermore, the current version of the compiler
expects static array allocation, although this is not a fundamental
limitation and we are working on supporting dynamic allocation. The
current OpenCL backend generates code that is optimized either for
CPU or for GPU and we are actively working on generating optimized
code for FPGAs.

\section{Conclusion}

We have developed a proof-of-concept compiler for OpenCL acceleration
and auto-parallelization of domain-specific legacy FORTRAN 77 scientific
code using whole-program analysis and source-to-source compilation.
We have validated the code transformation performance of the compiler
on the NIST FORTRAN78 test suite and a number of real-world codes;
the automatic parallelization component has been tested on a 2-D Shallow
Water model and on the Large Eddy Simulator for Urban Flows and produces
a complete OpenCL-enabled code base that is 20x faster on GPU than
the original code on CPU. Future work will focus on improving the
compiler to extract more parallelism from the original code and improve
the performance; and development of a complete FPGA back-end.

\section*{Acknowledgements}

The authors acknowledge the support of the EPSRC (EP/L00058X/1).

\section*{References}

\bibliographystyle{elsarticle-num}
\bibliography{paracfd2017-WV,rf4a-paper}

\end{document}